\documentclass[twocolumn,10pt]{asme2ej}

\usepackage{epsfig} 

\title{Heat Conduction in Nanostructured Materials Predicted by Phonon Bulk Mean Free Path Distribution}

\author{Giuseppe Romano
    \affiliation{
	Department of Materials Science and Engineering\\
	 Massachusetts Institute of Technology\\
	 77 Massachusetts Avenue, Cambridge (MA), 02139\\
           Email: romanog@mit.edu
    }	
}

\author{ Jeffrey C. Grossman
    \affiliation{ Department of Materials Science and Engineering\\
	 Massachusetts Institute of Technology\\
	 77 Massachusetts Avenue, Cambridge (MA), 02139\\
           Email: romanog@mit.edu
    }
}

\begin{document}

\maketitle    

\begin{abstract}
\it We develop a computational framework, based on the Boltzmann transport equation, with the ability to compute the thermal transport in nanostructured materials of any geometry using as the only input the bulk thermal conductivity accumulation function. The main advantage of our method is twofold. First, while the scattering times and dispersion curves are unknown for most materials, the phonon mean free path distribution can be directly obtained by experiments. As a consequence, a wider range of materials can be simulated than with a frequency-dependent approach. Second, when phonon dispersions are available from first principles calculations, our approach allows one to include easily the whole Brillouen zone in the calculations without discretizing the phonon frequencies  for all polarizations, reducing considerably the computational effort. Furthermore, after deriving the ballistic and diffusive limits of our model, we develop a multi-scale scheme that couples phonon transport across different scales, enabling efficient simulations of materials with wide phonon mean free path distributions. After we validate our model against the frequency-dependent approach, we apply the method to porous silicon membranes and find good agreement with experiments on mesoscale pores. By enabling the investigation of thermal transport in unexplored nanostructured materials, our method has potential to advance high-efficiency thermoelectric devices.

\end{abstract}

\section{Introduction}

Nanostructured materials have gained much attention for thermoelectric applications, thanks to their ability to effectively suppress phononic thermal conductivity (PTC)~\cite{majumdar2004thermoelectricity,dresselhaus2007new}. Very low thermal conductivities of thin films~\cite{venkatasubramanian2001thin}, nanowires~\cite{hochbaum2008enhanced} and porous materials~\cite{song2004thermal,yu2010reduction,tang2010holey,lee2008nanoporous} have been recently reported, reaching in some cases a suppression of two orders of magnitude with respect the bulk. However, understanding these phenomena is difficult, as classical size effects become important and diffusive theories fail when the characteristic material length (CML) becomes comparable to the average phonon mean free path (MFP)~\cite{chen2005nanoscale}. Phonon classical size effects have been long modeled by means of the Casimir model, that assumes that all phonons scatter diffusely at boundaries~\cite{Casimir1938495}. 
However, to effectively engineer thermal transport in arbitrary shapes and structures, a more accurate model of heat transport through a material is necessary. To this end, many studies employ the Boltzmann Transport Equation (BTE) to compute phonon transport~\cite{majumdar1993microscale,PhysRevB.57.14958,cahill2003nanoscale}. The simplest BTE model assumes that the phonon group velocities and scattering times are energy independent, \textit{i.e.} the medium is "gray". Although this so-called gray model has been useful for understanding trends in thermal transport in many nanostructured materials, it has poor predictive power especially for materials with wide MFP distributions. In some cases the MFP distribution spans several orders of magnitude. For example in Si, although the most commonly used value for the MFP is $\sim100~nm$, it has been predicted that half of the heat is carried by phonons whose MFP is greater than $1 \mu m$~\cite{eucken}.  Recent studies have addressed this challenge by including the full phonon distribution by successfully solving the frequency-dependent BTE (FD-BTE) for both transient and steady state cases~\cite{minnich2011quasiballistic,hsieh2012thermal,loy2013fast}. However, despite its  high accuracy, the FD-BTE suffers two major limitations. First, it requires knowledge of the actual phonon dispersion curves and the phonon-phonon relaxation times, which are generally unknown for new potential thermoelectric materials. The second limitation is related to computational efficiency: FD-BTE requires the discretization of the dispersion curves and thus dense sampling close to zero group velocity zones. Furthermore, as all polarizations have to be included, it becomes computationally prohibitive for materials with complex unit cells and phononic materials, where phonon dispersion curves have several branches. 

In this work, we provide a new form of the steady state BTE, with the capability to compute nanoscale heat transport using as input only the bulk MFP distribution, a quantity that can be directly obtained experimentally~\cite{minnich2011thermal,minnich2012determining}, while retaining the accuracy of the FD-BTE. Also, it allows one to include straightforwardly the whole Brillouin Zone (BZ) in the calculations, without relying on the isotropic BZ approximation, typically used in FD-BTE calculations. In addition, we couple our model, which we refer to as MFP-BTE, with the ballistic BTE and the Fourier model in order to include ballistic and diffusive effects, respectively, in a consistent and seamless manner. We apply our method to study three dimensional classical size effects in porous Si membranes and find good agreement with experiments on mesoscale size pores. As our method can be applied to any material whose MFP distribution is known, its applicability could enable a substantially broader range of simulations of thermal transport in nanomaterials. Furthermore, in contrast with the FD-BTE approach, the MFP-BTE method  is based on a discretization of the MFPs, and as a consequence, the computational time does not increase as the number of the phonon branches increases, enabling efficient simulations of materials with complex unit cells. 

\section{Phonon Boltzmann Transport Equation}

In analogy with photon radiation, nanoscale heat transport can be described by an intensity of phonons $I(\mathbf{r},\mathbf{s},\omega,p) = v(\omega,p) \hbar \omega f(\omega,p) D(\omega,p)/(4\pi)$, where $\mathbf{r}$ is the spatial coordinate, $\mathbf{s}$ is the direction of phonon transport within a unit solid angle, $\omega$ and $p$ are the phonon frequency and polarization, respectively, $v(\omega,p)$ is the magnitude of the phonon group velocity, $D(\omega,p)$ is the phonon density of states, $f(\omega,p) $ is the non-equilibrium phonon distribution and $\hbar$ is the reduced Plank's constant. In its original formulation, the FD-BTE describes the intensity of phonon $I$ under the relaxation time approximation~\cite{majumdar1993microscale}
\begin{equation}\label{eprt}
\frac{1}{v}\frac{\partial I}{\partial t} + \mathbf{s} \cdot \nabla I  = \frac{I^0(T)-I}{v\tau},
\end{equation}
where $\tau(\omega,p,T)$ is the scattering time and $I^0(T) = v(\omega,p) \hbar \omega f^0(\omega,T) D(\omega,p)/(4\pi)$ is an isotropic phonon intensity parametrized by the Bose-Einstein distribution $f^0(\omega,T) = \left[ exp(\hbar\omega / k_B T)-1  \right]^{-1}$ at a given local effective temperature $T(\mathbf{r})$~\cite{ziman2001electrons,mingo2014ab}.  We point out that $T(\mathbf{r})$ is not a thermodynamic temperature but rather should be considered as a measure of the average energy of phonons at a given point $\mathbf{r}$~\cite{PhysRevB.84.205331}. In order to compute the phonon thermal conductivity (PTC), we apply a difference of temperature $\Delta T$ across a simulation domain with length $L$ and calculate the thermal flux from the hot contact to the cold one. The  thermal flux $\mathbf{J}(\mathbf{r})$ is computed by 
\begin{equation}\label{eflux}
\mathbf{J}(\mathbf{r})= 4 \pi \sum_p  \int_0^{\omega_M^p} <I\mathbf{s}> d\omega, 
\end{equation}
where $< x> = \frac{1}{4\pi}\int_{4\pi}  x d\Omega$ is the angular average over the solid angle $4 \pi$ and $\omega_M^p$ is the frequency cut-off for a given polarization. In this work, we consider only the steady state BTE, \textit{i.e.} $\frac{\partial I}{\partial t} \approx 0$. The term $I^0(\mathbf{r},\mathbf{s},\omega,p)$ can be computed by applying the continuity equation for the heat flux, \textit{i.e.} $\nabla \cdot \mathbf{J}(\mathbf{r}) = 0$ to Eq.~\ref{eprt}, which yields 
\begin{equation}\label{econs}
\sum_{p}  \int_0^{\omega_M^p} \frac{I_0(T)}{v \tau} d\omega = \sum_p \int_0^{\omega_M^p} \frac{ < I >  }{v\tau}d\omega. 
\end{equation} 
We note that as both $I^0(\omega,p,T)$ and $I(\omega,p)$ are frequency dependent, knowledge of the phonon dispersion curves and scattering times is required.

\section{MFP dependent BTE}

In the following, starting from Eqs.~\ref{eprt}-\ref{econs} we develop a new version of the BTE where the only required input is the bulk MFP distribution, given by~\cite{yang2013mean}
\begin{equation}\label{acc}
K(\Lambda) = -\frac{\Lambda}{3} \sum_p  C_s v \left(\frac{d\Lambda}{d \omega}\right)^{-1}, 
\end{equation}


where $C_s(\omega)$ is the spectral heat capacity (\textit{i.e.} the volumetric heat capacity times the phonon density of states) and $\Lambda(\omega) = v(\omega)\tau(\omega)$ is the MFP. If the applied temperature gradient is small enough to assume that all the materials properties are constant throughout the simulation domain, we can define the variable
\begin{equation}\label{new}
\tilde{T} =  \frac{T_{\omega,p}-T_0}{\Delta T} = 4 \pi \frac{I-I_0(T_0)}{C_s v \Delta T},
\end{equation}
representing the departure from $T_0$ of the effective temperature $T_{\omega,p}$ associated to a given phonon frequency, polarization and direction, normalized by $\Delta T$. For simplicity,  in our simulations we use $\Delta T = 1$ and $T_0 = 0$. We note that a similar formulation is used in the low-variance deviatonal Monte Carlo approach, developed by Peraud \textit{et al}~\cite{PhysRevB.84.205331,APLNicholas}. Including the first order Taylor expansion~\cite{minnich2012determining} of $I^0(\omega,p,T) = I^0(\omega,p,T_0) + \frac{1}{4\pi}C_s(\omega,p)v(\omega,p)\left(T-T_0\right)$ in Eqs.~\ref{econs}, we have
\begin{eqnarray}\label{fdbte0}
\Lambda \mathbf{s} &\cdot& \nabla[I - I^0(T_0)] + I - I^0(T_0)  = \nonumber \\ &=&C_s v \left[\sum_p \int_0^{\omega_p'}\frac{C_s}{\tau}d \omega' \right]^{-1} \sum_{p'}\int_0^{\omega_M^p}  \frac{<I - I^0(T_0)>}{\Lambda}    d\omega', 
\end{eqnarray}
where we use the fact that $I^0(T_0)$ is isotropic and spatial independent.   
We note that Eq.~\ref{fdbte0} still needs the phonon scattering times and dispersion curves. If we use the new variable $\tilde{T}$ as given in Eq.~\ref{new}, Eq.~\ref{fdbte0} can be rearranged into
\begin{equation}\label{mfpbte}
\Lambda \mathbf{s} \cdot \nabla   \tilde{T} + \tilde{T} = \gamma \int_0^\infty   \frac{K}{\Lambda'^2}   <\tilde{T}>  d\Lambda',
\end{equation}
where $\gamma = \left[\sum_p \int_0^\infty  \frac{K}{\Lambda^2} d\Lambda \right]^{-1}$ is a material property, which for Si is $\gamma_{Si} = 2.2739\cdot 10^{-17} K/W$. The right hand side  of Eq.~\ref{mfpbte} is equal to $\frac{T(\mathbf{r})-T_0}{\Delta T}$, where $T(\mathbf{r})$ is the effective local temperature. Eq.~\ref{mfpbte} is solved iteratively and the first guess for $T(\mathbf{r})$ is given by a Fourier simulation~\cite{romano2011multiscale}. Further details regarding the implementation of the spatial domain and the solid angle can be found in~\cite{romano2012mesoscale,romano2011multiscale}.  Since Eq.~\ref{mfpbte} requires as input only the bulk MFP distribution, we will refer to it as the MFP dependent BTE or, in short, \textit{MFP-BTE}. 
\subsection{Boundary conditions}
According to~\cite{jeng2008modeling,hsieh2012thermal}, for the FD-BTE, periodic boundary conditions have to be applied to the departure of the phonon intensity from equilibrium. For example, if $\mathbf{P}$ is the periodicity vector, we have
\begin{equation}\label{bc}
 I(\mathbf{r + P},\mathbf{s},\omega,p) -I_0(\mathbf{r + P},\omega) = I(\mathbf{r},\mathbf{s},\omega,p) -I_0(\mathbf{r },\omega)
 \end{equation}
It is possible to show that in the case of MFP-BTE we simply need to apply $\Delta \tilde{T}=1$. Partially diffusive boundary conditions at a surface with normal $\mathbf{n}$ can be applied by imposing 
\begin{equation}\label{bc}
\tilde{T}(\mathbf{r},\mathbf{s})=(1-p)\frac{1}{\pi}\int_{\mathbf{s'}\cdot\mathbf{n}>0}\tilde{T}(\mathbf{r},\mathbf{s'})\mathbf{s'}\cdot\mathbf{n}d\Omega + p \tilde{T}(\mathbf{r},\mathbf{s_i})
\end{equation}
where $\mathbf{s_i}=\mathbf{s} - 2 |\mathbf{s}\cdot \mathbf{n}|\mathbf{n}$ is the specular direction related to the surface with normal $\mathbf{n}$ and $p$ is the specularity parameter, which depends on the surface roughness~\cite{ziman2001electrons}. 

\section{Effective thermal conductivity}

In the new variable formulation, the thermal flux can be obtained as $\mathbf{J} =3 \Delta T  \int_0^\infty \frac{K(\Lambda)}{\Lambda} <\tilde{T}>\mathbf{s}d\Lambda$. Once Eq.~\ref{mfpbte} converges, we can compute the PTC by combining Eq.~\ref{eflux} with Fourier's Law 
\begin{equation}\label{eff}
\kappa_{eff}=\frac{3L}{A} \int_{\Gamma}\int_0^\infty \frac{K(\Lambda)}{\Lambda} <\tilde{T} \mathbf{s} \cdot \mathbf{n}> d\Lambda dS ,
\end{equation}
where $\Gamma$ is either the cold or hot contact and $A$ is its area. In Eq.~\ref{eff}, we use $<I^0\mathbf{s}> = 0$, because $I^0$ is isotropic. We note that Eqs.~\ref{mfpbte}-\ref{eff} do not require the knowledge of $\Delta T$. It is useful at this point to define the phonon suppression function
\begin{equation}\label{supp}
S(\Lambda)=\frac{3L}{ \Lambda A}\int_{\Gamma} <\tilde{T}  \mathbf{s} \cdot \mathbf{n} >  dS ,
\end{equation}
which yields a simpler formula for the PTC $ \kappa_{eff} = \int_0^\infty K(\Lambda) S(\Lambda) d\Lambda$. If we denote the MFP distribution in the nanostructured material as $K^{nano}(\Lambda)$, the phonon suppression function can be formally defined by $S(\Lambda) = K^{nano}(\Lambda)/K(\Lambda)$. Although defined differently, $S(\Lambda)$ is essentially the same as the material independent \textit{boundary function} $D(\Lambda^{nano}/\Lambda) = \Lambda^{nano}/\Lambda$ introduced by Dames \textit{et al} in~\cite{yang2013mean}, where $\Lambda_{nano}$ is the MFP in the nanostructured material, except that within our formalism $S(\Lambda)$ may in general depends on the material. We note that the physical meaning of $S(\Lambda)$ is slightly different from the phonon suppression function defined in recently developed MFP distribution measurements techniques~\cite{minnich2011thermal,minnich2012determining}, where the suppression function refers to the suppression of heat flux due to the finite thermal length induced in the bulk material.

\subsection{Ballistic limit}
In order to derive the ballistic limit of the MFP-BTE, we define the Knusden number as $Kn = \Lambda/L$. For nanostructure with $Kn >> 1$, phonons travel mainly ballistically and their effective MFPs approach the characteristic length of the material. Within this regime, it is possible to show that the mode temperature distributions are MFP independent. Under this simplification,  Eq.~\ref{mfpbte} turns into 
\begin{equation}\label{rte}
  \Lambda \mathbf{s} \cdot \nabla \tilde{T}  + \tilde{T} = const, 
\end{equation}
which is the ballistic BTE~\cite{chen2001ballistic}. Eq.~\ref{rte} si computationally less expensive than the MFP-BTE, because phonons with different MFPs are decoupled from each other. 

\subsection{Diffusive limit}
Following a similar approach as~\cite{loy2013fast}, the diffusive limit of the MFP-BTE can be obtained by using the first spherical expansion of $\tilde{S}(\mathbf{r},\mathbf{s},\Lambda)$. Given a small perturbation $\mathbf{\Phi}(\mathbf {r})\cdot \mathbf{s}$ in the $\mathbf{s}$ direction, the inclusion of the spherical expansion of $\tilde{T}(\mathbf {r},\mathbf {s},\Lambda) = <\tilde{T}(\mathbf {r},\mathbf {s},\Lambda)> + \mathbf{\Phi}(\mathbf {r})\cdot \mathbf{s}$ into Eq.~\ref{mfpbte} leads to
\begin{equation}\label{sphere}
\tilde{T}= \gamma \int_0^\infty\frac{K}{\Lambda'^2}<\tilde{T}>d\Lambda' - \Lambda \nabla\cdot <\mathbf{s} \tilde{T}>.
\end{equation}
where $\nabla \cdot \mathbf{\phi}<<\Lambda \nabla\cdot <\mathbf{s} \tilde{T}>$ is used. The corresponding heat flux $\mathbf{J} = - K 3 \Delta T  \nabla <\tilde{T} >$, which is obtained by multiplying both sides of Eq.~\ref{sphere} by $\frac{3 \Delta T K}{\Lambda}\mathbf{s}$ and integrating over the solid angle, is now diffusive. The continuity equation for the thermal flux, derived by multiplying by $3 \Delta T K(\Lambda)/\Lambda$ both sides of Eq.~\ref{mfpbte} and computing an angular average, is given by
\begin{equation}\label{flux}
\nabla \cdot \mathbf{J} = \frac{3 \Delta T K}{\Lambda^2}\left(  \gamma \int_0^\infty   \frac{K}{\Lambda'^2}  <\tilde{T}>  d\Lambda' -  < \tilde{T}>  \right).
\end{equation}
The MFE in Eq.~\ref{mfe} is then obtained by combining Eq.~\ref{flux}-\ref{sphere} and dividing both sides by $3 \Delta T \frac{K}{\Lambda^2}$. The resulting equation 
\begin{equation}\label{mfe}
\Lambda^2\nabla^2 <\tilde{T}> - <\tilde{T}> =  \gamma \int_0^\infty\frac{K}{\Lambda'^2}<\tilde{T}>d\Lambda' 
\end{equation} 
is diffusive equation with a right hand side acting as an effective heat source and describes the energy balance among different phonon modes. We refer to Eq.~\ref{mfe} as the Modified Fourier Equation (MFE)~\cite{loy2013fast}. 

\section{Multiscale approach}

The solution of Eq.~\ref{mfpbte} requires the discretization of the simulation domain into a mesh whose characteristic size should be at least as small as the smallest MFP, making the phonon transport calculation in materials with wide MFP distributions computationally expensive. To solve this numerical issue we define a threshold in the Knusden number $Kn_{D}$, below which phonons are solved by means of the MFE. Involving directly $<\tilde{T}>$, Eq.~\ref{mfe} is computationally less intensive than the MFP-BTE, at the expense of neglecting phonon size effects~\cite{loy2013fast}.  
We also identify another threshold $Kn_B$, above which all phonons are solved by means of Eq.~\ref{rte}. As in the ballistic regime all phonon modes are decoupled from each other, and Eq.~\ref{rte} has to be solved only once.  By solving iteratively  Eqs.~\ref{mfpbte}-\ref{rte}-\ref{mfe}, we ensure energy conservation among all phonon modes, which are linked to each other through $T(\mathbf{r})$. The first guess for $T(\mathbf{r})$ is given by a Fourier's simulation~\cite{romano2011multiscale}. A general approach for choosing the transition points among different regimes is to start by the reasonable guess $Kn_D=Kn_B=1$. Then we push $Kn_D$ and $Kn_B$ toward lower and higher values, respectively, until convergence in the heat flux is reached. 

\section{Code validation}

In order to assess the accuracy of our model, we compare the MFP-BTE with the FD-BTE for two dimensional porous Si. Then, we ensure that the diffusive limit is fully recovered by comparing analytical data with results obtained with a macroscale domain.

\subsection{FD-BTE/MFP-BTE comparison}
\begin{figure}[htb]
\begin{center}
\includegraphics[width=8cm]{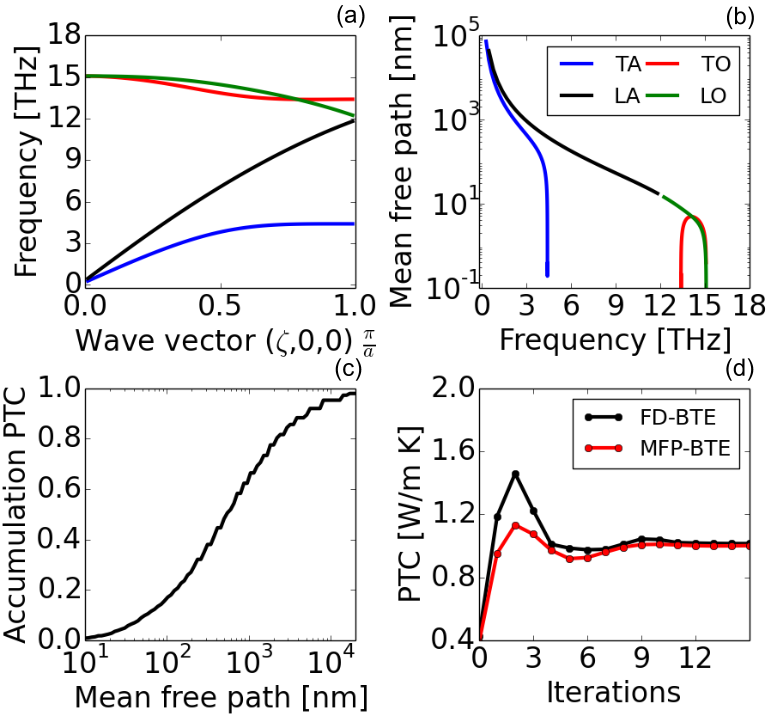}
\caption{a) Phonon dispersion along the \textit{001} direction computed by first-principles calculations b) The MFPs for different phonon polarizations c) The accumulation PTC at room temperature d) Comparison between the FD-BTE and the MFP-BTE method for a unit cell of size $L=10nm$. After few iterations, the two methods leads to the same thermal conductivity.}\label{fig:figure1}
\end{center}
\end{figure}

The FD-BTE requires the knowledge of the phonon dispersions and scattering times. Specifically for the FD-BTE/MFP-BTE comparison, the Brillouin zone is assumed isotropic and only the dispersion curves along $\textit{001}$ are considered~\cite{minnich2011quasiballistic}. The phonon frequencies, shown in~\ref{fig:figure1}-a,  are computed by means of first principles calculations by using the Quantum Espresso software package~\cite{giannozzi2009quantum}. The Umklapp and isotopic phonon scattering times are obtained by 
\begin{eqnarray}\label{scattering}
1/\tau_u &=& B\omega^2T exp(C/T)\\
1/\tau_i &=& D\omega^4
\end{eqnarray}
where $A, B, C$ and $D$ are parameters used to fit experimental data on bulk Si~\cite{minnich2011quasiballistic}. The total scattering time is simply obtained by using the  Matthiessen's rule $1/\tau = 1/\tau_u  +1/\tau_i$. The phonon MFPs for each polarizations are shown in~\ref{fig:figure1}-b whereas the accumulation PTC used for the MFP-BTE is reported in~\ref{fig:figure1}-c. We consider a unit cell with length $L=10nm$ containing one pore with diffusive walls. Periodic boundary conditions along the heat flux direction are applied and a fixed porosity $\phi = 0.25$ is considered. According to Fig.~\ref{fig:figure1}-d, the MFP-BTE produces the same results as the FD-BTE, revealing the equivalence between the two methods. We note that for the MFP-BTE we have to discretize only one function, \textit{i.e.} the bulk MFP distribution, whereas the FD-BTE requires the discretization of the phonon dispersion curves for each polarization, increasing the computational effort especially for complex unit cell materials, such as $Bi_2Te_3$. The computational efficiency of the MFP-BTE can be further increased if we use a power law interpolation scheme. In fact, $\tilde{S}(\Lambda)$ is typically a smooth function in $\Lambda$, and goes as $1/\Lambda$ toward the ballistic limit. On the other side, for the FD-BTE the phonon dispersion curves may require a dense interpolation scheme in those zones with small group velocity~\cite{minnich2011quasiballistic}.

\subsection{Diffusive limit of porous materials}
We now further validate our code by allowing the length of the unit cell reaching macroscopic scales. In Fig.~\ref{fig:figure2}-a, the normalized effective temperature computed by the MFP-BTE  for the $L = 100~nm$ case is reported. As we can see from Fig.~\ref{fig:figure2}-b, phonons mostly travel in the spaces between pores, due to collisions with pore boundaries.
\begin{figure}[htb]
\begin{center}
\includegraphics[width=8.5cm]{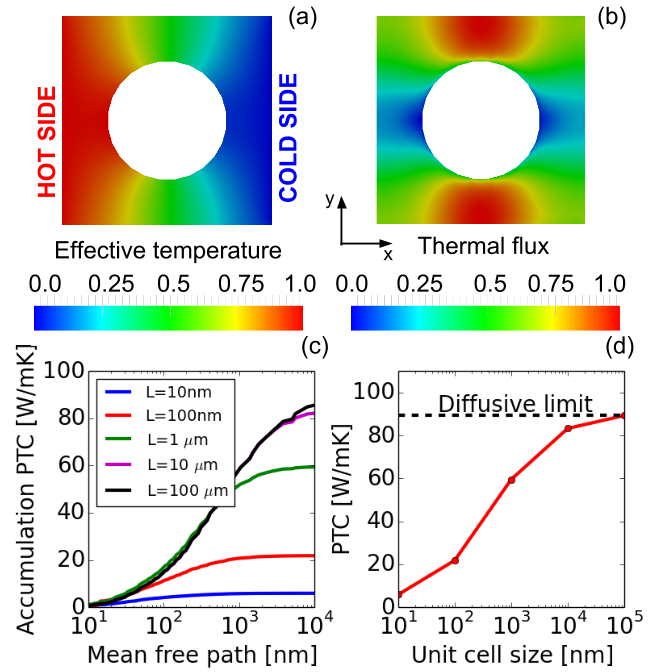}
\caption{ a) Effective temperature map for the $L = 100~nm$ case. The heat flux is ensured by applying a difference of temperature to the unit cell. b) Magnitude of thermal flux. Most of heat travels in the space between pores. c)  PTC accumulation function for different size of the unit cell $L$, ranging from the nanoscale to the macroscale. d) PTC versus the unit cell size. For macro scales, the effective thermal conductivity reaches the diffusive theory.}\label{fig:figure2}
\end{center}
\end{figure}
In Fig.~\ref{fig:figure2}-c the accumulation PTC $a^{nano}(\Lambda) = \int_0^{\Lambda} K(\Lambda') S(\Lambda') d\Lambda' $ is plotted for different length scales. According to Fig.~\ref{fig:figure2}-c, for $L=10~nm$ the PTC is $\kappa_{eff} = 5W/mK$ while for very large pores the thermal transport reaches the diffusive limit, predicted by Hashin and Shtrikman~\cite{hashin1962variational} $\kappa_{eff}/\kappa_{bulk} = (1-\phi)/(1+\phi)\approx 89.5 W/mK$. The optimum values for the transition Knusden numbers are $Kn_D = 0.1$ and $Kn_B = 10$. We have also performed calculations without using the multiscale scheme described above and similar results have been found.

\section{Application to porous silicon membrane}
In order to simulate three-dimensional realistic systems, we use as input the bulk MFP distribution computed by first-principles~\cite{esfarjani2011heat}, which takes into account the whole Brillouin zone. 

\subsection{Nanoscale porous membrane}
We calculate the thermal flux across the Si nanomesh studied in~\cite{yu2010reduction} with length $L=34~nm$, porosity $\phi = 0.173$ and height $H=22~nm$. In Fig.~\ref{fig:figure2}-a we plot the accumulation PTC. The measured value is about only $\kappa_{eff} = 2.85 W/mK$~\cite{yu2010reduction}, while the computed PTC is $~8 W/mK$. Besides unavoidable errors in measuring thermal conductivities at the nanoscale, this discrepancy could suggest the presence of phonon wave effects. However, according to~\cite{MinnichNM}, such a low experimental thermal conductivity can be justified by the presence of an amorphous layer along the pores wall, leading to a higher effective porosity. We remark that our work neglects wave effects hence the reduction of the PTC arises only from the scattering between phonons and the boundaries.  Phonon wave effects may dominate thermal transport at room temperature in certain systems such as those composed of thin films with periodic nanopillars~\cite{nanopillars}, where local resonances cause band flattening. Such effects and wave effects in general are beyond the scope of our work.

\subsection{Mesoscale porous membrane}
We now consider a mesoscale porous Si membrane with $L=4\mu m$ and height $H=4.49\mu m$, as reported experimentally in Ref.~\cite{song2004thermal}.  As shown in ~\ref{fig:figure2}-b, heat travels primarily in the spaces between pores along the direction of the imposed temperature gradient. The top and bottom surfaces are considered purely diffusive and act as an additional scattering source. This effect can be further understood if we examine the cut in the magnitude of thermal flux reported in Fig.~\ref{fig:figure2}-c. In fact, most of the thermal flux is concentrated in the middle of the sample, leading to a further reduction in the PTC. As the scattering with the top and bottom surfaces is not included in the MFE, the optimum threshold in the Knusden number is lower than the one used for the 2D case and turns out to be $K_D = 0.025$. The computed PTC is about $\kappa_{eff} = 56 W/mK$ while the experimental value is about $45 W/mK$. Considering the error in the measurements for thin films~\cite{jain2013phonon}, the two results are in good agreement with each other. We note that the best numerical estimation was $67 W/mK$, obtained by the MFP sampling method~\cite{jain2013phonon}. We remark that, although the simulation domain is in the mesoscale regime, the above results have been obtained with no input parameters, revealing the validity of our multiscale method. 
\begin{figure}[htb]
\begin{center}
\includegraphics[width=8cm]{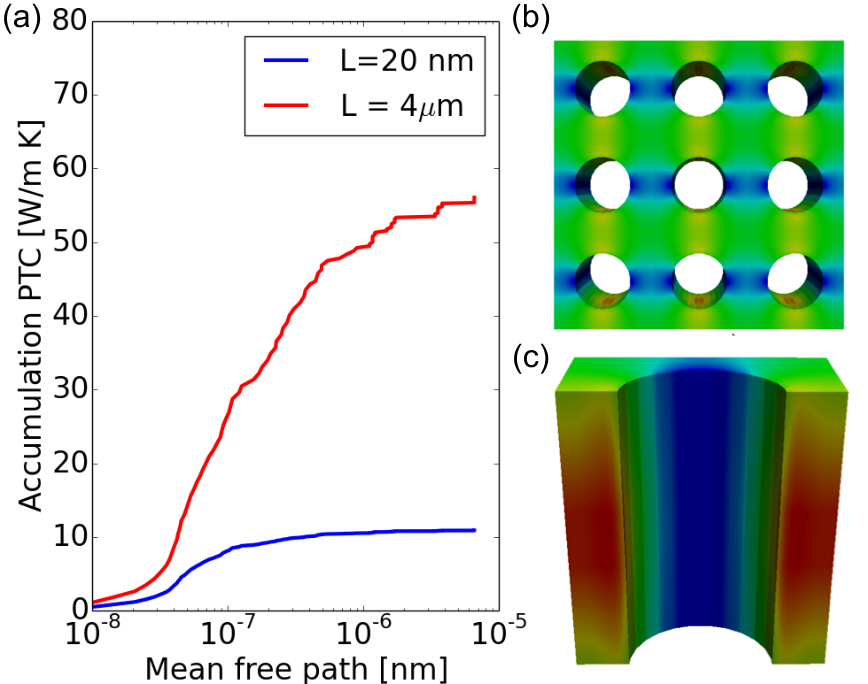}
\caption{a) the PTC accumulation function for a mesoscale pores Si membrane~\cite{song2004thermal} and Si nano mesh~\cite{yu2010reduction}. Phonon classical size effects strongly depend on the limiting dimension that is the smallest between the pore size and the sample thickness. b) The magnitude of the thermal flux for the periodic structure. Phonons travel between pores along the direction of the gradient of the temperature, due to phonon-pore scattering. c) A cut of magnitude of the thermal flux map. Most of the heat is concentrated toward the middle of the sample due to the diffuse scattering of phonons with the top and bottom surfaces}\label{fig:figure2}
\end{center}
\end{figure}

We remark that while the method has been applied to Si, its validity is general and any material can be modeled, as long as its bulk MFP distribution is  known, either experimentally or computed by first principles. 

\section{Conclusions}
In summary, we have developed an efficient method based on the BTE with the ability to compute the PTC without requiring prior knowledge of the phonon dispersion curve and three-phonon scattering times, using as input only the bulk MFP distribution, which can be directly obtained by experiments. In addition to its wide range of applicability, this method is more computationally efficient than the FD-BTE, in particular for materials with complex unit cells. Our results show good agreement with measurements on mesoscale porous Si membranes, showing the validity of the model across different length scales. 

\begin{acknowledgment}
We thank Keivan Esfarjani and Gang Chen for providing the first principles bulk thermal conductivity accumulation function for Si. We thank Yongjie Yu, Aldo Di Carlo and Alexie Kolpak for useful discussions.
\end{acknowledgment}

%
% ...

\bibliographystyle{asmems4}





\end{document}